\documentclass[12pt]{article}
\usepackage{amssymb,graphics,axodraw}
\newcommand{\newc}{\newcommand}
\newc{\beqa}{\begin{eqnarray}}
\newc{\eeqa}{\end{eqnarray}}
\newc{\beq}{\begin{equation}}
\newc{\eeq}{\end{equation}}
\newc{\bitem}{\begin{itemize}}
\newc{\eitem}{\end{itemize}}
\newc{\nonr}{\nonumber}
\newc{\hs}{\hskip 0.2cm}
\newc{\ra}{\rightarrow}
\newc{\ep}{\epsilon}
\newc{\tri}{\triangle}
\newc{\fpr}{\frac{1}{\sqrt{2\pi R}}}

\begin{document}
\pagestyle{empty} \vskip -2cm

%
\begin{flushright}
\today
\end{flushright}
\begin{center}{\bf \large
Lepton Universality, Rare Decays and Split Fermions }

\vskip 1cm

{\large
We-Fu Chang$^{a}$, I-Lin Ho$^{a,b}$
 and John N. Ng$^{a}$
}\\
{\em $^a$TRIUMF Theory Group, 4004 Wesbrook Mall, Vancouver, B.C. V6T 2A3 CA }\\
{\em $^b$Physics Department, National Tsing-Hua University,\\
Hsinchu 30043, Taiwan, R.O.C.}\\

\end{center}
%
\vskip 1.0cm
\begin{abstract}
We investigate the constraint on the split fermions in extra
dimensions by considering the universality of $W$ leptonic decays
$W\rightarrow l_i \nu_i$, the charged  lepton decays $l_i \ra
l_j\nu_i\bar{\nu}_j$, and  the lepton flavor violating process $l_i\ra
\bar{l}_j l_k l_h$ where
$l_i= e,\mu$ or $\tau$. For the Standard Model (SM) background of
$W\ra l_i \nu_i$, we extended the one loop quantum correction to
include effects of order $m_l^2/M_W^2$ and the Higgs mass
dependence. We find that in general the split fermion scenarios give rise to
a 4D effective Yukawa matrix of the Kaluza-Klein Higgs bosons is
misaligned with respect to the fermion mass matrix. This holds true
also for gauge bosons as well. This leads to
decays of $l_i\ra \bar{l_j}l_k l_h$ at tree level and muonium antimuonium conversion.
Interestingly the leptonic
universality of $W$ boson decays are not affected at this level.
\vskip 0.5cm
\centerline{PACS numbers:  11.25.Mj,  12.60.-i,  11.10.Kk}
\end{abstract}

\newpage
\pagestyle{plain}
\section{Introduction}
Recently new avenues of exploring physics beyond the Standard
Model(SM) have opened up by assuming that there exists large extra
dimensions beyond the four we are familiar with
\cite{ADD},\cite{Anton},\cite{JL}. The earlier investigations have the graviton and
 possibly SM
gauge  singlet  particles such as the right-handed neutrinos
are allowed to propagate in the extra
dimensions whereas the particles that have SM charges
are confined to a 4 dimensional hypersurface known as the
TeV or the SM brane. This picture can also provide a natural geometrical
understanding of the hierarchy of fermion masses by postulating
that the chiral fermions of the SM are localized at different
points in the extra dimensions \cite{AS}; i.e. they are split from
each other. By the same token different families of fermions also
occupy different points in bulk space. The localization of a
chiral fermion is represented by a Gaussian wave function in the
extra dimension $y$. The mass of a fermion is generated via a five
dimensional Yukawa term. In four dimensions, after integrating out
$y$,  a small Yukawa coupling arises due to the  small overlap of
the wave functions of the left- and right- handed components of a
fermion. In this way a hierarchy in the effective 4D Yukawa
couplings is obtained without invoking new symmetries. A detail
model for the observed quark and lepton masses in terms of their
displacements in $y$ has been given in \cite{MS}.
Besides offering a new vista on the Yukawa coupling hierarchy this
scenario also points to a novel way of looking at the question of
gauge coupling universality. Historically, the branching ratio
(Br) of $\pi \ra e\nu$/$\pi \ra \mu \nu$ provided  the crucial
evidence that the charged weak current couples with the same
strength to the first two lepton families. This universality study
has since been extended to leptonic $\tau$ decays and also to the
leptonic branching ratios of the $W$ boson. These are cornerstones
that support the SM and they are very accurately predicted in the
SM. An example is the Br$(W\ra l_i \nu/W\ra l_j \nu)= 1 +
O(\alpha)$ where $l_i=e,\mu,\mathrm {or} \;\tau$. In the SM the
deviation from unity is a function of lepton masses and the lepton
energy cut used in a given experiment \cite{univ}. The dependence
on the unknown Higgs boson mass is very weak. As a by-product of
our investigation we will give the complete 1-loop SM result. We feel that it
is very important to examine how proposed new physics will altered
these predictions.

To see more quantitatively how these various purely leptonic reactions can be used
to probe the split fermion scenario we study the simplest
model with the minimal SM in 5D. The chiral fermions are confined at different positions
in the extra dimension. The exact mechanism of localization is not
important to our study and we shall leave it open. The $SU(2) \times U(1)$ gauge bosons
and the Higgs doublet are allowed to propagate in the full 5D bulk. The model is compactified
in a $S_1/Z_2$ orbifold of radius $R$ with the appropriate boundary conditions so as to preserve the
successes of the 4D SM. An interesting  generic feature of the split fermions scenario is the
existence of effective flavor changing neutral currents which we
shall demonstrate are related to the separation between two chiral
fermions belonging to different families.
This is first noticed in \cite{Quiros}
for the KK gauge bosons in the model. We extend this to both the neutral and the
charged KK Higgs bosons. In this paper we
concentrate on the issue of lepton flavor violation (LFV)
interactions partly because they involve less theoretical
uncertainties. We will also concentrate on tree level predictions of charged lepton
decays.
We will not discuss the many issues related to neutrino mass in the bulk world scenario even though
they are
very interesting. Neutrino masses can be made natural within
the framework of extra dimensions \cite{bnu} by introducing one \cite{bnu1} or more
$\nu_R$ 
and phenomenological studies of the properties of
bulk neutrinos can be found in \cite{bnup} and references therein. Localizing the
right-handed bulk neutrinos at different points in the extra dimension is performed in
 \cite{Tait} and \cite{Baren}. The latter reference also discussed
lepton flavor violation in a general way which is different from our treatment. After
acknowledging this we shall assume that neutrino masses are due to yet unknown
4D new physics and its phenomenology is beyond the scope of this paper.

In the quark sector an extensive numerical study of the quark mass matrices and their mixings
is given in \cite{MS} whereas
the issue of CP
violation has been investigated in \cite{Branco}. Furthermore
supersymmetry has also been incorporated in this scenario in \cite{Tait}. A related but
different scheme using multilocation is given in \cite{das}

This paper is organized as followed. In Sec 2 we give the details of the 5D SM model and obtain
the 4D effective interactions after integrating out the extra dimension.
The Feynman rules for the 4D interactions are summarized in the Appendix. Sec.3 gives the phenomenology
of both normal and rare decays of charged leptons. In particular the decay of
$\mu$ to 3e provides the strongest constrain to the parameters of the model.
 In Sec 4  we  compare the tests of universality
using leptons and on shell
W-boson decay. They are  shown to be complementary and will be an important task
for the Large Hadron Collider currently under construction. Finally we give our
conclusions in Sec 5.

\section{5D SM Model with Split Leptons}

The model we employed is the 5D SM similar to that introduced in
Ref.\cite{Quiros} augmented by the distributions of chiral
fermions located at different points in the extra dimension,$y$. It is a crucial assumption that
 the left-handed ($L$) lepton doublet is separated from
the right-handed ($R$) lepton. For the minimal matter content of
the SM ignoring neutrino mass there are six independent locations  $y^a_i$ where we use $i,j,k$
for  family indices and $a, b\in\{L, R\}$ stand for
chiralities. One of these can be chosen as the origin.
 The 4D effective theory is obtained by
compactifying the bulk fields  on a $S_1/Z_2$ orbifold where $S_1$
is a circle define by $-\pi R\leq y\leq\pi R$.
Strictly speaking  $R$ is a free parameter and is bounded by experiments.
We shall assume that  $R \lesssim(300 \mathrm {GeV})^{-1}$ for the sake of phenomenological interest.
Then we implement the idea that chiral fermions can be trapped at
topological domain wall in such a setting \cite{trapf} and also
 at different locations \cite{AS}, \cite{DS}. The zero mode of a fermion
is chiral and is given a narrow Gaussian distribution in $y$.  We adopt a
universal Gaussian width $\sigma$ for all the fermions.
We use the notation that the coordinates in Minkowski space is denoted by
$x^{\mu},\mu=\{0\cdots 3\}$ and in bulk space by
 $x^M,M=\{0\cdots 3,y\}$.
Also the fifth Dirac matrix is chosen to be $\gamma^y=i\gamma_5$.

 The 5D SM Lagrangian is given by
\beqa
{\cal L}_5&=&-\frac14 F^{MN}F_{MN}-\frac14
G^{(c),MN}G^{(c)}_{MN}
+\overline{L'}(x,y) i\gamma^M D_M L'(x,y)\nonr\\
&&+\left(D_M \Phi(x,y)\right)^\dag \left(D^M \Phi(x,y)\right)
-\kappa  R\left(\mid\Phi(x,y) \mid^2- \frac{v_b^3}{2}\right)^2\nonr\\
&&-\sqrt{2\pi R}f_{ij}\overline{L'}_i(x,y)\Phi(x,y)E'_j(x,y)
+ h.c.+\cdots
\label {eqn:L5}
\eeqa
where $L'$ and $E'$ are respectively
the $SU(2)$ doublet and singlet lepton fields.  $\Phi$ is the bulk Higgs field and
$v_b$ is its vacuum expectation value (VEV). For simplicity, we
take a universal Yukawa coupling $f$ which is of order one for
$f_{ij}$. Also, as in the SM,
\beqa
D_M&=&\partial_M -i g_2 \sqrt{2\pi R} \frac{\tau^{c}}{2}B_M^{c}
- i g_1 \sqrt{2\pi R} \frac{Y}{2}A_M, \nonr\\
F_{MN}&=& \partial_M A_N - \partial_N A_M, \nonr\\
G^{(c)}_{MN}&=& \partial_M B^{c}_N - \partial_N B^{c}_M
+\sqrt{2\pi R}g_2 \epsilon^{cde}B_M^{d} B_N^{e}
\eeqa

and $A$, $B$  stand for the $U(1)$ hyper charge and $SU(2)$
gauge fields. In this convention, $Q=T_3+Y/2$. Note that the mass dimensions
of various quantities are: $[\Psi]\!=\!2$,
$[\Phi]\!=\!\frac32$, $[g_1]\!=\![g_2]\!=0$,
$[\kappa]\!=\! 0$  and $[f]\!=\! 0$.

 In our study  we can ignore KK excitations of the fermions
but will  keep the  KK excitations of  Higgs and gauge bosons . For $\sigma\ll
R$, the chiral zero mode of a fermion field $\Psi^a_i$ located at
$y^a_i$ can be normalized to \beq \Psi^a_i(x,y)\sim {1 \over
\pi^{\frac14}\sigma^{\frac12} }\Psi^a_i(x)
e^{-\frac{(y-y^a_i)^2}{2\sigma^2}}. \eeq The  product of two
fermion fields can be approximately replaced by \beq
\overline{\Psi}^a_i(x,y)\Psi^b_j(x,y) \sim
\exp\left(-\frac{(\triangle_{ij}^{ab})^2}{4\sigma^2}\right)
\delta(y-\bar{y}^{ab}_{ij})\bar{\Psi}^a_i(x)\Psi^b_j(x) \eeq where
$\bar{y}^{ab}_{ij}=(y^a_i+y^b_j)/2$ is their average positions and
$\triangle^{ab}_{ij}=y^a_i-y^b_j$. It is known \cite{acd} that the
5D SM where all fields propagate in the full bulk there is conservation
of KK number in the 4D effective theory due to momentum conservation.
It is  interesting that when a Gaussian profile is given
to the fermion filed the usual KK number conservation now is
replaced by a suppression factor
\[
\cos\frac{n y_i}{R}e^{-\frac{n^2\sigma^2}{4R^2}}\;\;.
\]
This is because
\beq
\int^{\pi R}_{-\pi R} dy \cos\frac{n y}{R}
{1\over \pi^{1/2}\sigma}e^{-{(y-y_i)^2 \over \sigma^2} }
\cong \cos\frac{n y_i}{R} e^{-\frac{n^2\sigma^2}{4R^2}}.
\eeq
This is understood because that the Gaussian localized wave function serves as the
fifth momentum `reservoir' that compensates the momentum carried by
bulk gauge boson and thus maintains the conservation of momentum. Now it is possible
to have vertices with only one non-zero KK mode. Since we expect $\sigma \ll R$ the
exponential factor is almost unity.

The fermion and gauge boson masses are generated by the VEV
of bulk Higgs which we write as
\beq
\Phi(x,y)=
\left(\begin{array}{c} h^+(x,y) \\\frac{1}{\sqrt{2}}\left( v_b^{\frac32} +
\phi^0(x,y)\right)
\end{array}\right).
\eeq
The scalar field is taken to be even under $Z_2$ and the KK decomposition is given by
\beq
\phi^0(x,y)=\frac{1}{\sqrt{2 \pi
R}}\left(\phi^0_0(x)+ \sqrt{2}\sum_{n=1}^\infty
\phi^0_n(x)\cos\frac{n y}{R}\right).
\eeq
The zero mode $\phi^0_0(x)=h^0_0 +i\chi_0$ is
identified as the SM Higgs boson, $h_0^0$, and its Goldstone partner $\chi_0^0$. The KK tower
also contains a real and imaginary part given by $\phi^0_n= h_n^0 +i\chi_n^0$.
 A similar expression holds for
the gauge fields but their fifth component are assigned to be odd
under $Z_2$, so as to prevent the presence of the unwanted zero modes at the orbifold fix point.
This leads explicitly to the following expansion for gauge fields
\beqa
A^{\mu}(x,y)=\fpr \left( A_0^{\mu}(x)+\sqrt{2}\sum_{n=1}^{\infty}A_n^{\mu}(x)\cos\frac{n y}{R}\right), \nonr \\
A^4(x,y)=\frac{1}{\sqrt{\pi R}} \left(\sum_{n=1}^{\infty}A^4_n(x)\sin\frac{n
y}{R}\right).
\eeqa

As in 4D gauge theory one has to fix a gauge in a given calculation. The
5D generalization of covariant gauge
fixing Lagrangians are
\beqa
{\cal L}_{GF}&=& -\frac{1}{2\alpha} (\partial_M P^M )^2
         -\frac{1}{\xi}\left|\partial_M W^{+,M}-i \xi {g_5
         \cos\theta_W
         v_b^{3/2}\over 2}
         h^+\right|^2\nonr\\
       &&  -\frac{1}{2\eta}\left(\partial_M Z^M -\eta {g_5 v_b^{3/2}\over 2} \chi^0\right)^2
\eeqa
 where $\chi^0=\mbox{Im}(\phi^0)$ is the pseudoscalar would be Goldstone boson, and
$\alpha, \xi, \eta$ are the gauge parameters for the photon ($P$), W and Z bosons respectively.
Also  $g_5=\sqrt{2\pi R }\sqrt{g_1^2+g_2^2}\equiv\sqrt{2\pi R }g $.
Combining  ${\cal L}_5$ and ${\cal L}_{GF}$ and integrating  over $y$,
we get the 4D effective Lagrangian.
One finds that the usual tree level SM relations still held,
\beqa
g=\sqrt{g_1^2+g_2^2},\hskip1cm e=g_2\sin\theta_W,\hskip1cm
\tan\theta_W=\frac{g_1}{g_2},\nonr\\
M_Z= {g \sqrt{2\pi R}v_b^{3/2} \over 2},\hskip1cm M_W= M_Z \cos\theta_W.\nonr
\eeqa
It is straightforward to obtain the $n^{\mathrm th}$-KK gauge boson masses. They are:
$M_{\gamma,n}^2=n^2/R^2$;
 $M_{W,n}^2 = M_W^2+ n^2/R^2$; $M_{Z,n}^2 = M_Z^2 + n^2/R^2$.
 For the Higgs bosons one has $M_H^2=\kappa R v_b^3$ and
$M_{h^0,n}^2= M_H^2+n^2/R^2$.
In this model, the  fifth component of gauge fields behave like spin-0 particles
and their masses are
$M^2_{A^4,n}= n^2/(\alpha R^2)$, $M^2_{Z^4,n}=M_Z^2+n^2/(\eta R^2)$
and $M^2_{W^4,n}=M_W^2+n^2/(\xi R^2)$. They couple to the SM gauge bosons
on the brane but not couple to the brane fermions through gauge interaction.
 Another way to see the absence of couplings between the fifth components and
 brane fermions  is  that the interaction
$\bar{\Psi}A^4\Psi$ is odd under the orbifolding $Z_2$ parity.

The Higgs sector is same as in the SM,
\beq
-\sqrt{2\pi R} f \left[\bar{\nu}_{L i}h^+
+\frac{1}{\sqrt2}\left(v_b^{3/2}+h^0+i\chi^0\right)\bar{e}_{Li}
\right]e_{Rj}+ H.c.
\eeq
The masses of the would-be-Goldstone bosons and their KK partners  are: $M^2_{\chi^0,n}=\eta M_Z^2+n^2/R^2$
and $M^2_{h^+,n}=\xi M_W^2+n^2/R^2$.
In the 4D effective Lagrangian the following mixing terms also appear
\beqa
&&\frac{n}{R}\frac{1-\alpha}{\alpha}(\partial^\mu P_n^4)P_{n\mu}
+\frac{n}{R}\frac{1-\eta}{\eta}(\partial^\mu Z_n^4)Z_{n\mu}\nonr\\
&&+ \frac{n}{R}\frac{1-\xi}{\xi}\left(\partial^\mu W_n^{4+} W^-_{n\mu}
+\partial^\mu W_n^{4-} W^+_{n\mu}\right).
\label{eq:mix5}
\eeqa
Clearly by choosing  the Feynman gauge, $\alpha=\xi=\eta=1$, one can eliminate the mixing terms.
This is the most convenient gauge for calculating physical processes.
We summarize the Feynman rules we employ in the Appendix.

The effective 4D Yukawa coupling for charged leptons is
 \beq {\cal L}_Y = - f e^{-\frac{(\triangle^{LR}_{ij})^2}{4\sigma^2}}\overline{L'_i}(x) \left[\sqrt{\pi
R}v_b^{\frac32}+ \frac{\phi^0_0}{\sqrt{2}} + \sum_{n=1} \phi^0_n
\cos\frac{n\overline{y^{LR}_{ij}}}{R}e^{-\frac{n^2\sigma^2}{4R^2}}\right] E'_j(x) + H.c. \eeq
The mass matrix  is readily seen to be
\beq \label{mmatrix} {\cal
M}_{ij}= f\sqrt{\pi R} v_b^{\frac32}
e^{-\frac{(\triangle^{LR}_{ij})^2}{4\sigma^2}}=
f\frac{\sqrt{2}M_W}{g_2}e^{-\frac{(\triangle^{LR}_{ij})^2}{4\sigma^2}}
\simeq f m_t
e^{-\frac{(\triangle^{LR}_{ij})^2}{4\sigma^2}}.
\eeq
Temporarily suppressing family indices, the mass matrix  is diagonalized
by a bi-unitary transformation as follows
\beq
 M_{diag}=V^{L\dag} M V^R,\hskip5mm L'(x)=V^L L(x), \hskip5mm E'(x)= V^R E(x)
\eeq
where  $L(x)$ and $E(x)$ are mass eigenstates.

It is easy to see that this diagonalization also rotates away the
off-diagonal coupling of fermions to the Higgs zero mode; i.e. the
SM Higgs to fermion couplings remain flavor diagonal. Furthermore,
the SM gauge bosons fermion couplings are also flavor diagonal.
However, the Higgs KK modes will couple different mass eigenstate
fermions as displayed in
\beq
{\cal L}= -\bar{L}_i\left(m_i+ {g_2
m_i\over 2M_W} \phi^0_0\right)E_i - \sum_{n=1}
\lambda^{LR}_{ij,n}\bar{L}_{i}E_{j}\phi^0_n +h.c.
\eeq
where
\beq
\lambda^{LR}_{ij,n}= f V^{L*}_{k i}
\exp\left({-\frac{(\triangle_{kl}^{LR})^2}{4\sigma^2}}\right)
\cos\frac{n\bar{y}^{LR}_{kl}}{R} V^R_{lj}e^{-\frac{n^2\sigma^2}{4R^2}}.
 \eeq
 In other words one {\it cannot} simultaneously diagonalize the fermion mass matrix
and the Yukawa matrix of the KK Higgs-fermion couplings  due to
the presence of the cosine terms.
Similar flavor nondiagonal couplings are induced for the KK excitations of the $W$ and $Z$ bosons as well.
After some algebra, the effective 4D charged current Lagrangian can be cast in the form
\beq
\label{eq:CCL}
{\cal L}_{eff}^{CC}=g_2\overline{L_i}\left[ \gamma^\mu \frac{\tau^+}{\sqrt{2}}%
\left(\delta_{ij}W_{0,\mu}^+ +\sum_{n=1} W_{n,\mu}^+(x)U_{ij}^{L(n)}\right)\right]L_j + h.c.
\eeq
and the neutral Lagrangian is
\beqa
\label{eq:NCL}
{\cal L}_{eff}^{NC}&=& {g_2 g_L\over \cos\theta_W} \overline{L_i}\left[ \gamma^\mu %
\left(\delta_{ij}Z_{0,\mu} +\sum_{n=1}Z_{n,\mu}(x)U_{ij}^{L(n)}\right)\right]L_j\nonr\\
&+& {g_2 g_R\over \cos\theta_W}\overline{E_i}\left[ \gamma^\mu %
\left(\delta_{ij}Z_{0,\mu} +\sum_{n=1}Z_{n,\mu}(x)U_{ij}^{R(n)}\right)\right]E_j \nonr \\
&+& h.c
\eeqa
where  $g_{L/R}=T_{3,L/R}-Q_{L/R} \sin^2\theta_W$.
And
\beqa
\label{Umat}
U^{L(n)}_{ij}&=&\sqrt{2}\sum_{k=1}^3 V^{L*}_{ki} \cos\frac{n y_k^L}{R} V^L_{kj}e^{-\frac{n^2\sigma^2}{4R^2}},\nonr\\
U^{R(n)}_{ij}&=& \sqrt{2}\sum_{k=1}^3 V^{R*}_{ki} \cos\frac{n y_k^R}{R}
V^R_{kj}e^{-\frac{n^2\sigma^2}{4R^2}}.
\eeqa
The very same mixings $U^{L(n)}_{ij}$ and $U^{R(n)}_{ij}$ are also associate with
the KK photon.

Clearly the KK excitations of the photon, the $Z$ boson, and the Higgs boson will all induce tree level
lepton flavor violation processes. We will discuss their contributions in detail in the next section.

\section{ Constraint on the fermion locations}

Equipped with the Feynman rules given we can proceed to discuss the phenomenology of charged lepton
decays. Consider first classic case of muon decay into an electron and a pair of neutrinos.
At the tree level the SM has only one amplitude involving $W$ boson will
now have contributions also from its KK excitations. The split fermion scenario also adds
the contributions from KK excitations of the $Z$ (see Eq.\ref{eq:NCL})
 and the charged Higgs boson. Also a sum over the neutrinos in the
final state is taken. Similar modifications to the usual discussions of rare decays
also occurs and they are systematically presented in the following subsections.

\subsection{Lepton Universality}
In terms of mass eigenstate the effective 4D charge current
interaction is given by Eq.(\ref{eq:CCL}).
 Since we only have one bulk Higgs
field  there is no mixing between physical $W$ boson, which is the
zero mode, and its KK excitation. Therefore, it is universally
coupled to the lepton families.
 We conclude that lepton universality tested by ratios of
the leptonic width of the $W$ boson will remain at the SM values at the lowest level.

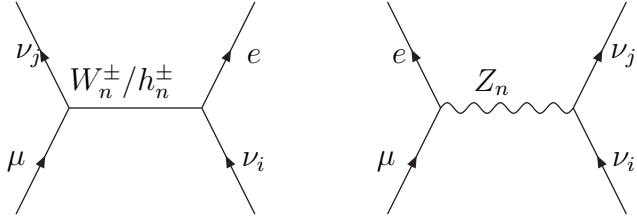
\begin{figure}[htbp]\begin{center}
\begin{picture}(250,100)(0,0)
\ArrowLine(10,10)(30,50) \Text(10,30)[c]{$\mu$}
\ArrowLine(30,50)(10,90) \Text(15,70)[c]{$\nu_j$}
\Line(30,50)(80,50) \Text(50,60)[c]{$W_n^\pm / h_n^\pm$}
\ArrowLine(100,10)(80,50) \Text(100,30)[c]{$\nu_i$}
\ArrowLine(80,50)(100,90) \Text(100,70)[c]{$e$}

\ArrowLine(150,10)(170,50) \Text(150,30)[c]{$\mu$}
\ArrowLine(170,50)(150,90) \Text(155,70)[c]{$e$}
\Photon(170,50)(220,50){2}{5} \Text(190,60)[c]{$Z_n$}
\ArrowLine(240,10)(220,50) \Text(240,30)[c]{$\nu_i$}
\ArrowLine(220,50)(240,90) \Text(240,70)[c]{$\nu_j$}
\end{picture}\end{center}
\caption{ Contribution of KK excitations to $\mu$ decay  }
\end{figure}

 On the other hand for the  classic decay of
$\mu\rightarrow e \nu\bar{\nu}$, where all the possible virtual  KK modes also
participate, information on  the $y$-dependence can be gleamed.
Neglecting the electron mass  we get
\beqa
\label{mudecay}
-{\cal M}
&=& \left({g_2 \over \sqrt{2}}\right)^2\left[{ \delta_{i1}\delta_{j2} \over M_W^2}
+\sum_{i,j,n} { U^{L(n)}_{i1}U^{L(n)*}_{j2}\over
M_{W,n}^2}\right]
(\bar{\nu}_j \gamma^\alpha_L
\mu)(\bar{e}\gamma_{L,\alpha}\nu_i)\nonr\\
&-& \sum_{i,j,n} {2 \lambda^{LR}_{i2,n}\lambda^{LR*}_{j1,n}\over
M_{h^+,n}^2}
(\bar{\nu}_j \mu)(\bar{e} \nu_i)\nonr\\
&+&  g_L^\nu g_L^e\left({g_2 \over \cos\theta }\right)^2 \sum_{i,j,n} { U^{L(n)}_{12}U^{L(n)*}_{ji}\over M_{Z,n}^2}
(\bar{\nu}_j\gamma^\alpha_L \nu_i)(\bar{e}\gamma_{L,\alpha}\mu)\nonr\\
&+&  g_L^\nu g_R^e\left({g_2 \over \cos\theta }\right)^2 \sum_{i,j,n} { U^{R(n)}_{12}U^{L(n)*}_{ji}\over M_{Z,n}^2}
(\bar{\nu}_j\gamma^\alpha_L \nu_i)(\bar{e}\gamma_{R,\alpha}\mu)\nonr\\
&\equiv& \left({ g^2_2 \over 2M_W^2 }\right)\left\{( 1+a_1 R^2M_W^2)
(\bar{\nu} \gamma^\alpha_L \mu)(\bar{e}\gamma_{L,\alpha}\nu)\right. \nonr \\
&&- a_2 M_W^2R^2
(\bar{\nu}\mu_R)(\bar{e}_R\nu )\nonr\\
&&\left.+ a_3  M_W^2 R^2 (\bar{\nu} \gamma^\alpha_L\mu)(\bar{e}\gamma_{L,\alpha} \nu)
-2 a_4 M_W^2 R^2\left(\bar{\nu} \mu_R\right)\left(\bar{e}_R
\nu\right)\right\}.
\eeqa
The Fierz transformation have been used to get the last expression.
The coefficients $a_{i,j,k}$ are the result of summing  over all neutrino species. Explicitly,
\beqa
a_1\sim \sum_{i,j,n=1} { U^{L(n)}_{j2}U^{L(n)*}_{1i} \over
n^2},\hskip2mm
a_2\sim \sum_{i,j,n=1} { 4 \lambda^{LR}_{j2,n}\lambda^{LR*}_{i1,n} \over g^2_2 n^2},\nonr\\
a_3\sim \left({2 g_L^\nu g_L^e \over \cos^2\theta }\right)
\sum_{i,j,n=1} { U^{L(n)}_{12}U^{L(n)*}_{ij} \over n^2},\hskip2mm
a_4\sim\left({2 g_L^\nu g_R^e \over \cos^2\theta }\right)
\sum_{i,j,n=1} { U^{R(n)}_{12}U^{L(n)*}_{ij} \over n^2 }.
\eeqa
We define the process dependent Fermi constant $G_{\mu}$ as
\beq
{4 G_{\mu}\over \sqrt{2} }={g^2_2 \over 2M_W^2 }
\left[\left(1+R^2M_W^2(a_1+a_3)\right)^2
+\left({a_2+2a_4\over 2}\right)^2R^4M_W^4\right]^{\frac12}.
\eeq
The square bracket gives the modification to the SM Fermi coupling
constant, $G_{SM,F}=\sqrt{2}g^2_2/8M_W^2$ and also generalizes the usual KK result
\cite{KKG}.
Eq.(\ref{mudecay}) reveals that the Michel parameters $\rho$ and $\delta$ will have the SM
value of 3/4. This is easily seen in the
charge retention mode. On the other hand we have
\[
\eta \simeq -\frac{( a_2+2a_4) R^2M_W^2}{2}.
\]
Thus, we expect a deviation from the SM value of $\eta=0$.
 This in turn leads to the following prediction for
 the partial decay width of a charged lepton  into a  purely leptonic
 channel $l_i\ra l_j+\bar{\nu} \nu$ in this model :
\beqa
\label{eq:width}
\Gamma(l_i\ra l_j+\bar{\nu} \nu)=
{m_i^5 G_{ij}^2 \over 192 \pi^3 }
\left\{1-8\alpha^2_{ij}+8\alpha^6_{ij}-\alpha^8_{ij}-24\alpha^4_{ij}\ln\alpha_{ij}\right.\nonr\\
\left.+4 \eta_{ij} \alpha_{ij}(1+9\alpha_{ij}^2-9\alpha_{ij}^4-\alpha_{ij}^6)%
+48\eta_{ij} \alpha_{ij}^3(1+\alpha_{ij}^2)\ln\alpha_{ij} \right\}
\eeqa
where $\alpha_{ij}=\frac{m_j}{m_i}$, $G_{ij}$ represents the specific Fermi constant
 for the process $l_i\ra l_j+\bar{\nu} \nu$.
Expand in  powers of $M_W^2 R^2$ and keeping the lowest order we obtain
\beqa
\label{eq:HLCLD}
\Gamma(l_i\ra l_j+\bar{\nu} \nu)
\sim
{m_i^5 G_{SM,F}^2 \over 192 \pi^3}\left[\frac{}{}1-8\alpha^2_{ij}+8\alpha^6_{ij}-\alpha^8_{ij}
-24\alpha^4_{ij}\ln\alpha_{ij}\right.\nonr\\
\left.+2(a_1+a_3)M_W^2
R^2-2\alpha_{ij}(a_2+2a_4)M_W^2R^2\frac{}{}\right].
\eeqa
Now  consider the tauon partial decay width ratio
$\Gamma(\tau\ra \mu\bar{\nu} \nu)/\Gamma(\tau\ra e \bar{\nu}
\nu)$,
using the current experimental limit \cite{PDG} we find
\beq
2M_W^2R^2\left[a^{\tau\mu}_1+a^{\tau\mu}_3-a^{\tau e}_1-a^{\tau e}_3
-\frac{m_\mu}{m_\tau}\left(a^{\tau\mu}_2+2a^{\tau\mu}_4\right)
\right] \leq .003 \eeq
which is a constraint on chiral fermion geography.

 For a simple illustration, we use the example of a  diagonal charged lepton mass
 matrix as given in \cite{MS}. Actually this kind of mass matrix
 is  unnatural in this setup. However it can be achieved
 by pairing the left-handed and right-handed leptons in the
 same generation into a cluster which is well separated from the
 other two generations' clusters. Choosing this setting, the flavor
 violating gauge coupling are highly suppressed and the Yukawa
 coupling of KK Higgs is proportional to charged lepton mass, so
 \[
 a_2,\hskip2mm a_3,\hskip2mm a_4 \sim 0.
 \]
The universality breaking now is solely  from the $y-$dependent
couplings of KK $W$.
Since the Gaussian width $\sigma$ is much smaller than the radius
$R$, the exponential suppression factor can be ignore to a good
approximation. The series can be summed \cite{ssum} and keeping the
lowest order in $y/(\pi R)\ll 1$ we get
 \beqa
a_1^{\tau \mu}\sim 2\sum_{n=1}\frac{1}{n^2}
\cos\frac{n y_\tau^L}{R}\cos\frac{n y_\mu^L}{R}
= \frac{\pi^2}{6}\left[2-3\frac{| y_\tau^L+y_\mu^L|}{R\pi}
 -3{|y_\tau^{L}-y_\mu^{L}| \over R\pi}\right]+{\cal O}\left(\frac{y^2}{\pi^2 R^2}\right)\\
a_1^{\tau e}\sim 2\sum_{n=1}\frac{1}{n^2}
\cos\frac{n y_\tau^L}{R}\cos\frac{n y_e^L}{R}
= \frac{\pi^2}{6}\left[2-3\frac{|y_\tau^L+y_e^L|}{R\pi}
-3{|y_\tau^{L}-y_e^{L}| \over R\pi}\right]+{\cal O}\left(\frac{y^2}{\pi^2 R^2}\right)
 \eeqa
We have the limit
\beq
\label{eq:tau_U}
M_W^2 R \pi \left(|y_\tau^L+y_e^L|+|y_\tau^{L}-y_e^{L}|
- | y_\tau^L+y_\mu^L| - |y_\tau^{L}-y_\mu^{L}|
\right)<0.003.
\eeq
If we choose  $y_\tau^L=0$  and $ R= 300\mbox{GeV}^{-1}$ then we have
\[
 {|y_e^L|-|y_\mu^L|\over  R}< 6.6\times 10^{-3} \left({R^{-1}\over
 300 \mbox{GeV}}\right)^2
 \]
which give limits on  the separation between different generations.

 We emphasize that even for this extreme
 case the fermion location dependence still breaks the charged
 lepton universality. It is a generic feature of this model.
 For more general non-diagonal mass matrices such as that studied in \cite{Ross}\cite{Jap}
 the KK Z and $h^+$ will also contribute to the breaking of
 universality. We note that the charged lepton matrix of \cite{Jap} cannot
be easily incorporated in the split fermion scenario although it has other success.

 We see from the above that flavor violating neutral currents are generic in this scenario.
In next section we will discuss the constraint from flavor violating reactions.
\subsection{$\mu \mathrm{\ and\ }  \tau$ to three charged leptons}
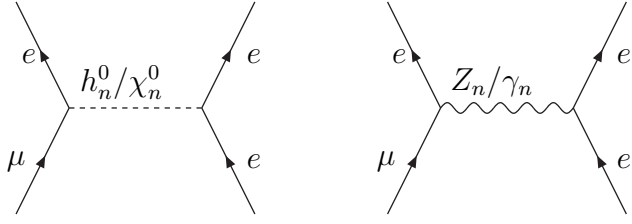
\begin{figure}[htbp] \begin{center}
\begin{picture}(250,90)(0,20)
\ArrowLine(10,10)(30,50) \Text(10,30)[c]{$\mu$}
\ArrowLine(30,50)(10,90) \Text(15,70)[c]{$e$}
\DashLine(30,50)(80,50){2} \Text(50,60)[c]{$h^0_n/ \chi^0_n$}
\ArrowLine(100,10)(80,50) \Text(100,30)[c]{$e$}
\ArrowLine(80,50)(100,90) \Text(100,70)[c]{$e$}

\ArrowLine(150,10)(170,50) \Text(150,30)[c]{$\mu$}
\ArrowLine(170,50)(150,90) \Text(155,70)[c]{$e$}
\Photon(170,50)(220,50){2}{5} \Text(190,60)[c]{$Z_n/\gamma_n$}
\ArrowLine(240,10)(220,50) \Text(240,30)[c]{$e$}
\ArrowLine(220,50)(240,90) \Text(240,70)[c]{$e$}
\end{picture}\end{center}\caption{ Diagrams for $\mu$ decays into
3 electrons.}\label{fig:mu3e} \end{figure}

Due to the existence  of flavor violating interactions in the gauge and the Higgs sectors,
the following processes l: $\mu \ra 3e$, $ \tau \ra
3e$, $\tau \ra \mu e e$, $\tau\ra \mu \mu e$ and $\tau \ra 3 \mu$  will be induced by virtual
 KK Z, photon, scalar and pseudoscalar boson exchange at tree level as shown in Fig.\ref{fig:mu3e}.
The present upper branch ratio limit for the muon is around
$10^{-12}$ and for the $\tau$  is about $10^{-6}$ \cite{PDG}.

The low energy effective Lagrangian for $\mu\ra 3e$ is easily calculated to be
\beqa
 \sum_{n=1}\left\{
  { M_{\chi^0_n}^2- M_{h^0_n}^2 \over M_{h^0_n}^2  M_{\chi^0_n}^2}
 \lambda^{LR}_{ee,n} \lambda^{LR}_{e\mu,n}(\bar{e}_L e_R)(\bar{e}_L \mu_R)
+ { M_{\chi^0_n}^2+ M_{h^0_n}^2 \over M_{h^0_n}^2  M_{\chi^0_n}^2}
 \lambda^{LR*}_{ee,n} \lambda^{LR}_{e\mu,n}(\bar{e}_R e_L)(\bar{e}_L\mu_R)\right.\nonr\\
+ { M_{\chi^0_n}^2- M_{h^0_n}^2 \over M_{h^0_n}^2  M_{\chi^0_n}^2}
 \lambda^{LR*}_{ee,n} \lambda^{LR*}_{\mu e,n}(\bar{e}_R e_L)(\bar{e}_R \mu_L)
+ { M_{\chi^0_n}^2+ M_{h^0_n}^2 \over M_{h^0_n}^2  M_{\chi^0_n}^2}
 \lambda^{LR}_{ee,n} \lambda^{LR*}_{\mu e,n}(\bar{e}_L e_R)(\bar{e}_R\mu_L)\nonr\\
-{g^2_2\over \cos^2\theta_W}{1\over M_{Z,n}^2}%
\left[\bar{e}\gamma^\mu(g_L (U^{L,n }_{ee})^*\hat{L}+g_R (U^{R,n}_{ee})^*\hat{R})e\right]%
\left[\bar{e}\gamma_\mu(g_L U^{L,n}_{e\mu}\hat{L}+g_R
U^{R,n}_{e\mu}\hat{R})\mu\right]\nonr\\
\left.-{e^2\over M^2_{\gamma,n}}%
\left[\bar{e}\gamma^\mu( (U^{L,n}_{ee})^*\hat{L}+ (U^{R,n}_{ee})^*\hat{R})e\right]%
\left[\bar{e}\gamma_\mu( U^{L,n}_{e\mu}\hat{L}+ U^{R,n}_{e\mu}\hat{R})\mu\right]\right\}+ h.c.
\eeqa
the sign difference in front of pseudoscalar term is
due to the nature of its imaginary coupling.
One can see that the terms $(\bar{e}_Le_R)(\bar{e}_L\mu_R)$ and
 $(\bar{e}_Re_L)(\bar{e}_R\mu_L)$
are almost vanishing  because  KK scalar and KK pseudoscalar  are nearly degenerated if we
assume that the SM Higgs mass is not too much heavier than the $W$ boson mass. For the high
KK states we can ignore these masses.

To a good approximation we can  neglect final state lepton masses
and obtain the branching ratio:
\beq
\label{etaijk}
B(\mu\ra3e)={\Gamma(\mu\ra 3e)\over
 \Gamma(\mu \ra e \bar{\nu}\nu)}
\sim2M_W^4R^4\left[  s_{LL}^2+ s_{RR}^2+2s_{LR}^2+2s_{RL}^2 + 4v_{RR}^2+ 4v_{LL}^2 \right]
\eeq
where
{
\setcounter{enumi}{\value{equation}}
\addtocounter{enumi}{1}
\setcounter{equation}{0}
\renewcommand{\theequation}{\theenumi.\alph{equation}}
\beqa
v_{LL}=\left(\sin^2\theta_W +\frac{g_L^2}{\cos^2\theta_W}\right)\sum_{n=1}
\frac{1}{n^2}(U^{L(n)}_{ee})^*U^{L(n)}_{e\mu}, \\
v_{RR}=\left(\sin^2\theta_W +\frac{g_R^2}{\cos^2\theta_W}\right)\sum_{ n=1}
\frac{1}{n^2}(U^{R(n)}_{ee})^*U^{R(n)}_{e\mu}, \\
s_{LL}=- \sum_{n=1} { \lambda^{LR*}_{\mu e,n} \lambda^{LR}_{ee,n}\over g^2_2 n^2},\hskip5mm
s_{RR}=-\sum_{n=1} { \lambda^{LR}_{e\mu,n} \lambda^{LR*}_{ee,n}\over g^2_2 n^2},\\
s_{LR}=-\left(\sin^2\theta_W +\frac{g_L g_R}{\cos^2\theta_W}\right)\sum_{ n=1}
\frac{1}{n^2}(U^{L(n)}_{ee})^*U^{R(n)}_{e\mu}, \\
s_{RL}=-\left(\sin^2\theta_W +\frac{g_L g_R}{\cos^2\theta_W}\right)\sum_{n=1}
\frac{1}{n^2}(U^{R(n)}_{ee})^*U^{L(n)}_{e\mu}.
\eeqa
\setcounter{equation}{\value{enumi}}
\addtocounter{equation}{-1}
}
Assuming a universal Yukawa coupling,  $f=1$, and again we take
the nearly diagonal mass matrix as an example. Then the flavor violating coupling will only
appear in Higgs sector. It predicts
$ v_{LL},\hskip1mm v_{RR},\hskip1mm s_{LR},\hskip1mm s_{RL} \sim 0$
and
\beqa
\label{eqn:etaterms}
\addtocounter{equation}{1}
\lambda^{LR}_{ee,n}
&\sim& {g_2 m_e \over \sqrt{2}M_W}\cos\frac{n
\bar{y}_{ee}^{LR}}{R}, \nonr\\
\lambda^{LR}_{e\mu,n}&\sim& \exp\left(-{(\tri^{LR}_{e\mu})^2 \over 4\sigma^2}\right)\cos\frac{n
\bar{y}_{e\mu}^{LR}}{R}.
\eeqa
So in this case,
\beq
s_{LL}\sim -\frac{\pi^2}{6}{m_e \over \sqrt{2}M_W g_2} \exp(-{(\tri^{LR}_{21})^2\over 4
\sigma^2}),\hskip3mm
s_{RR}\sim -\frac{\pi^2}{6}{m_e \over \sqrt{2}M_W g_2} \exp(-{(\tri^{LR}_{12})^2\over 4
\sigma^2}).
\eeq

Taking $R^{-1}\sim 300$ GeV, a value within the bound universal bulk models we obtain
\beq
B(\mu\ra3e)\sim  2.6\times 10^{-12}
\left[ \exp\left(-\frac{(\tri^{LR}_{12})^2}{2\sigma^2}\right)+
\exp\left(-\frac{(\tri^{RL}_{12})^2}{2\sigma^2}\right) \right ] \left({300 \mbox{GeV} \over
R^{-1}}\right)^4
\eeq
which is not far from the experimental limit.
The branching is also a very sensitive probe of the compactification radius $R$.
Similar expressions can be derived for $\tau$ flavor violating decays which
is also much lower  than current experimental limits.

For $B(\tau^-\ra e^+e^-\mu^-)$
and $B(\tau^-\ra e^-\mu^-\mu^+)$ the formula is slightly different since there no identical
particles in the final states.
\beq
B(\tau^-\ra e^+e^-\mu^-)\sim(2M_W^2R^2)^2[s_{LL}^2+s_{RR}^2+s_{LR}^2+s_{RL}^2+v_{RR}^2+v_{LL}^2]
\eeq
with replaced flavor indices $\mu e\ra \tau e$ and $ee\ra\mu e$ in Eq.(\ref{eqn:etaterms}).

We want to stress again we have used a special diagonal mass matrix to indicate what can be
learned. In general, the flavor violating process if observed will set constraint on how far the
leptons should be away from each other and Eqs.(\ref{etaijk}) and (\ref{eqn:etaterms}) should be
used. But than numerical method is
needed to solve them.

Without a theoretical understanding of the dynamics that determines the locations of the
fermions, we do a brute force numerical study.
A Monte Carlo program is composed to scan the possible mass matrices which can be
accommodated in this model and also satisfy the observed charged lepton
masses. Then the rotation matrices  $V^L$ and $V^R$ are calculated for
each set of the lepton locations. These in turn are used
 to calculate all the coupling vertices  of KK bosons. Of the half million of mass
matrices we scanned many contained off diagonal elements. However, only a
small fraction ($<5\%$) of these  can pass the experimental
bound of $B(\mu\ra3e)<10^{-12}$. For those satisfying the rare decays limit we use
their parameters to calculate the  Michel parameter $\eta$.
Setting the ratio $(\sigma/R)=1/50$, the numerical results give
$|\eta| \leq 3 \times 10^{-6}(300\mbox{GeV}/ R^{-1})^2$.
This will present a formidable experimental challenge. There is a direct
correlation between $\eta$ and $\mu \ra 3e$ since the KK $Z$ and Higgs exchanges
that will affect $\eta$ also contribute to the rare decay process. The stringent
experimental constrain for the rare decay mode sets the upper limit for $\eta$.
We note in passing that the situation for $\tau$ decays
is more hopeful since the constraint from its rare decays are less severe.

\subsection{Muonium-Antimuonium Conversion}

\begin{figure}[htbp]\begin{center}
\begin{picture}(400,100)(0,0)
\GOval(15,45)(25,5)(0){0.5} \Text(10,85)[c]{$M$}
\GOval(85,45)(25,5)(0){0.5} \Text(90,85)[c]{$\bar{M}$}
\ArrowLine(15,20)(50,25) \Text(30,15)[c]{$e$}
\ArrowLine(50,25)(85,20) \Text(70,15)[c]{$\mu$}
\ArrowLine(50,65)(15,70) \Text(30,75)[c]{$\mu$}
\ArrowLine(85,70)(50,65) \Text(70,75)[c]{$e$}
\DashLine(50,25)(50,65){3} \Text(60,50)[c]{$h^0/\chi^0$}
\Text(5,10)[c]{$(a)$}

\GOval(115,45)(25,5)(0){0.5} \Text(110,85)[c]{$M$}
\GOval(185,45)(25,5)(0){0.5} \Text(190,85)[c]{$\bar{M}$}
\ArrowLine(115,20)(150,25) \Text(130,15)[c]{$e$}
\ArrowLine(150,25)(185,20) \Text(170,15)[c]{$\mu$}
\ArrowLine(150,65)(115,70) \Text(130,75)[c]{$\mu$}
\ArrowLine(185,70)(150,65) \Text(170,75)[c]{$e$}
\Photon(150,25)(150,65){3}{4} \Text(165,45)[c]{$Z/\gamma$}
\Text(105,10)[c]{$(b)$}

\GOval(215,45)(25,5)(0){0.5} \Text(210,85)[c]{$M$}
\GOval(285,45)(25,5)(0){0.5} \Text(290,85)[c]{$\bar{M}$}
\ArrowLine(215,20)(235,45) \Text(230,25)[c]{$e$}
\ArrowLine(265,45)(285,20) \Text(270,25)[c]{$\mu$}
\ArrowLine(235,45)(215,70) \Text(230,65)[c]{$\mu$}
\ArrowLine(285,70)(265,45) \Text(270,65)[c]{$e$}
\DashLine(235,45)(265,45){3} \Text(255,55)[c]{$h^0/\chi^0$}
\Text(205,10)[c]{$(c)$}

\GOval(315,45)(25,5)(0){0.5} \Text(310,85)[c]{$M$}
\GOval(385,45)(25,5)(0){0.5} \Text(390,85)[c]{$\bar{M}$}
\ArrowLine(315,20)(335,45) \Text(330,25)[c]{$e$}
\ArrowLine(365,45)(385,20) \Text(370,25)[c]{$\mu$}
\ArrowLine(335,45)(315,70) \Text(330,65)[c]{$\mu$}
\ArrowLine(385,70)(365,45) \Text(370,65)[c]{$e$}
\Photon(335,45)(365,45){3}{4} \Text(355,55)[c]{$Z/\gamma$}
\Text(305,10)[c]{$(d)$}

\end{picture}\end{center}\caption{ Diagrams of Muonium-Antimuonium Conversion } \end{figure}
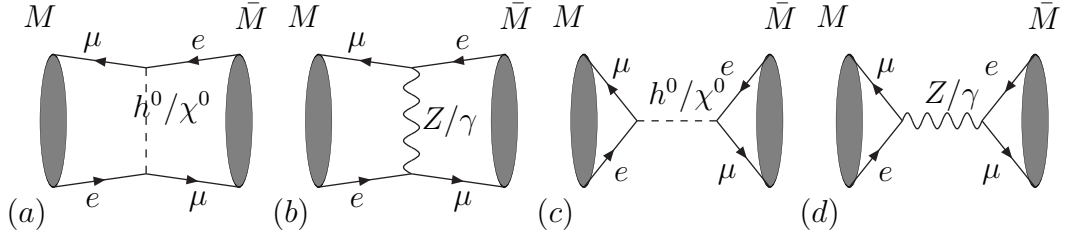

Unlike the rare decays  discussed before, here both $(V\pm A)^2-$ and $(V\pm A)(V\mp A)-$ type lepton number violating
interactions are present at the same time. The effective Hamiltonian for the conversion is given by
\cite{MAMC},
\[
{\cal H}=\frac{{\cal G}}{\sqrt{2}}(\bar{\mu}\gamma^\mu(1\pm\gamma^5)e)
(\bar{\mu}\gamma_\mu(1\pm\gamma^5)e)
+\frac{{\cal F}}{\sqrt{2}}(\bar{\mu}\gamma^\mu(1\pm\gamma^5)e)
(\bar{\mu}\gamma_\mu(1\mp\gamma^5)e)
\].
The muonium-antimuonium transition matrix elements involve different hyperfine states listed
below:
\beqa
\tri_{(1,\pm1)}&=&<\bar{M}_{1,\pm1}|{\cal H}|M_{1,\pm1}>={16 \over \sqrt{2}\pi
a_0^3}\left({\cal G}-\frac14{\cal F}\right),\nonr\\
\tri_{(1,0)}&=&<\bar{M}_{1,0}|{\cal H}|M_{1,0}>={16 \over \sqrt{2}\pi
a_0^3}\left({\cal G}-\frac14{\cal F}\right),\nonr\\
\tri_{(0,0)}&=&<\bar{M}_{0,0}|{\cal H}|M_{0,0}>={16 \over \sqrt{2}\pi
a_0^3}\left({\cal G}+\frac34{\cal F}\right),\nonr
\eeqa
where $a_0$ is the Bohr radius of the muonium $(m_r \alpha)^{-1}$
with $m_r^{-1}=m_\mu^{-1}+m_e^{-1}$.
Assuming that each state is produced with equal probability initially the integrated probability  of a muonium to
antimuonium conversion is
\beqa
P_{M\bar{M}}&=&64^3\left({2 \pi^2 \alpha^3\over G_F m_\mu^2}\right)^2 \left({m_e\over
m_\mu}\right)^6 \left({{\cal G}^2+\frac{3}{16}{\cal F}^2\over
G_F^2}\right)\\
&\simeq& 2.5 \times 10^{-5}\left({{\cal G}^2+\frac{3}{16}{\cal F}^2\over
G_F^2}\right).
\eeqa
The current experimental limit for this process is
$P_{M\bar{M}}\leq 8.3\times10^{-11}$\cite{MAMC_exp} which
implies
\beq
\sqrt{{\cal G}^2+\frac{3}{16}{\cal F}^2 }<
3.0\times 10^{-3} G_F \hskip3mm ( 90\% \mbox{C.L.}).
\eeq
 Taking diagonal mass matrix as an example again, the flavor
 violating interaction is mediated only by KK Higgs. Assume that $R\sim 300
 \mbox{GeV}^{-1}$ the above limit will translate into the following bound:
 \beq
 {(\tri^{LR}_{12})^2 +(\tri^{RL}_{12})^2 \over \sigma^2}> 13.8.
 \eeq

\section{ $W$ boson  universality in the SM}
As we have seen previously that the LFV mechanism predicts that
universality holds in $W$ boson decays but not $\mu$ or $\tau$
decays. Thus, it is important to establish the SM values for these
processes. We calculate the $W$ boson branching ratios to 1-loop
order in the on-shell scheme and used the unitary gauge which was
done in \cite{univ}. It is well known the width of $W\ra l \nu$ is
infrared finite only after including the radiative mode $W\ra l\nu
\gamma$ \cite{KLN}. After a laborious calculation we find the
leptonic decay width including undetected photon is
\beqa
{ \Gamma
\over \Gamma^{0}}=
\left[1+\frac{\alpha}{2\pi}\left\{2\left(2+{1+\beta\over 1-\beta}\ln\beta\right)%
\left(\ln\frac{M_W}{2\triangle E_l}+2\ln(1-\beta)\right) \right.\right.\nonr\\
\left.\left.-\frac32(1-\beta)\ln\beta +f_H\beta \right\}\right]
\eeqa
where $\beta\equiv m_l^2/M_W^2$ and $\Gamma^{0}=g_2^2 M_W^2(2+\beta)(1-\beta)/6$ is the lowest order width.
 The quantity $\triangle E_l$ is the finite energy resolution of the charged
lepton and is determined by a given experiment. $f_H$ is a
complicate function dependent on the Higgs mass. The exact form of
$f_H$ is not very illuminating and to a good approximation it is
\beq f_H\cong7.72+0.78\ln\frac{M_W^2}{M_H^2}. \eeq Numerically
the  values of $f_H$ are  $\{7.36, 6.32, 5.84, 5.34\}$
corresponding to Higgs masses of $M_H=\{ 110, 180, 250, 400\}$ GeV
respectively. Assuming that energy resolutions are the same for
all charged leptons the $W\rightarrow \nu e, \nu\mu,\nu\tau$ decay
width ratio is $1: 1.067 : 1.103$ for $\triangle E= 2$ GeV and $1:
1.038 : 1.057$ for $\triangle E= 5$ GeV. With the expected large
number of $W$ bosons to be produced at the LHC\cite{LHC} we can
expect this prediction to be tested in the near future.

\section{Conclusion}
We have shown that in the split fermion scenario with a bulk Higgs
boson it is not possible to diagonalize simultaneously the lepton
mass matrix and the Yukawa matrix of the KK Higgs modes. This complements
the FCNC arising from KK excitations of the $Z$ boson and the photon. This
leads to interesting new mechanism for rare $\mu$ and $\tau$
decays at the tree level without affecting lepton universality as probe by $W$ boson
decays.  On the other hand leptonic universality as probed by
$\tau$ lepton decays is altered by the virtual KK gauge boson and KK Higgs
exchanges. This gives a upper limit on the separation of different
families of leptons. In contrast rare LFV effects if seen are to
be understood as measuring the relative distances of a left-handed
fermion of one family to the right-handed fermion of a different
family in the extra dimension. If no signals are found in the next
round of experiments they give a lower bound on the fermion
separations. Similarly the fermion masses sets the relative
distances between fermions of opposite chiralities in the same
family.

Another interesting difference the split fermion scenario as compare
to usual extra dimension models is the non-universality of the KK gauge
bosons coupling the electron, muon and $\tau$. In other words when
a KK gauge boson, for instance the $n=1$ KK photon, is produced at high energies
 we expect its decay widths into
electrons, muons and taus to be different since the $U$ matrices of Eq.(\ref{Umat})
are sensitive to the family indices. This will lead to an apparent charge violation of
electric charge universality.

The above considerations can easily be extended to the quark
sector. The universality test of pion leptonic decay will set a
limit of ( see Eq.(\ref{eq:HLCLD} ) )
 \beq
 M_W^2R^2\sum_{n=1}\frac{1}{n^2}\mbox{Re}
 \left[ U^{L(n)}_{ud}\left( U^{L(n)*}_{11}-U^{L(n)*}_{22}\right )\right] \lesssim 10^{-7}
 \eeq
 assuming that it is dominated by flavor conserving KK $W$
 exchanging and $U^{L(n)}_{ud}$ is an obvious generalization to
the quark sector. In general we do not expect universality to hold
when comparing effective  charged current strengths as
measured in semileptonic versus leptonic experiments. This opens up
a new way of looking at the Cabibbo universality which we leave for future
considerations.

The split fermion scenario suffers from a number of drawbacks. On the theoretical
side there is a lack of understanding of the dynamics that determines the different
chiral fermion locations and the form of the wavefunctions. Phenomenologically there
are many parameters all of which has to be fixed by experiments.
 Despite all of these flaws some general distinguishing features
emerged. Among them is non-universality in both charged and neutral currents sectors
as probed by experiments done on the brane. On the other hand, the level
at which these effects are expected depends on the details of the model.

Our study has shown the importance of low energy precision tests
in covering the parameter space for these models. While it is too
early to do complete phenomenological analysis of even the minimal
model due to the scarcity of data at the same time we feel that
more studies involving similar rare processes are crucial. They
are complementary to direct collider searches for the KK
excitations of the SM particles.

 This work is supported in part by the Natural Science and
Engineering Council of Canada. We wish to thank Professor D. Chang
for the kind hospitality at the National Center of Theoretical
Science where this work was completed.

\newpage
\appendix
\section{Feynman Rules}
\setcounter{equation}{0}
\renewcommand{\theequation}{A.\arabic{equation}}
Combining  all pieces discussed in Sec 2 together, the relevant terms of the 5D SM are:
\beqa
{\cal L}_5+{\cal L}_{GF}=&-&\frac14 (\partial_M P_N-\partial_N P_M)^2-\frac14 (\partial_M Z_N-\partial_N
Z_M)^2\nonr\\
&-&\frac12\left(\partial_M W^+_N-\partial_N W^+_M\right)\left(\partial^M W^{-,N}-\partial^N
W^{-,M}\right)\nonr\\
&+&\frac12(\partial^M \phi^0)(\partial_M
\phi^{0*})+(\partial^M h^+)(\partial_M h^-)\nonr\\
&+&\frac14 (g v_0 c)^2 W_M^+W^{-,M}+ \frac18 (g v_0)^2  Z_M Z^{M}\nonr\\
&-&\frac{1}{\xi}(\partial^M W_M^+)(\partial^N W_N^-)
-\xi({g v_0 c\over 2})^2 h^+h^-\nonr\\
&-&\frac{1}{2\eta}(\partial^M Z_M)^2-\frac{\eta}{2}\left(\frac{g v_0}{2}\right)^2(\chi^0)^2
-\frac{1}{2\alpha}(\partial^M P_M)^2 +\cdots
\eeqa
where $P$ represents photon, $\phi^0= h^0 +i \chi^0$ and
$v_0=\sqrt{2\pi R}v_b^{\frac32}$.

Employing the KK decomposition, imposing the appropriate boundary conditions and
integrating  over $y$, we get the 4D effective Lagrangian
\beqa
&&{\cal L}_4= -\frac14 \sum_{n=0} P^{\mu\nu}_nP_{n,\mu\nu}
+\frac12 \sum_{n=1}\left[(\partial^\mu P^4_n)^2+\frac{n^2}{R^2}(P_n)^2-2\frac{n}{R}(\partial^\mu
P^4_n)P_{n,\mu}\right]\nonr\\
&&-\frac{1}{2\alpha}\sum_{n=0}\left[(\partial_\mu P^\mu_n)^2+2\frac{n}{R}\partial_\mu P^\mu_n
P_n^4
+\frac{n^2}{R^2}(P_n^4)^2\right]\nonr\\
&&-\frac14 \sum_{n=0} Z^{\mu\nu}_n Z_{n,\mu\nu}
+\frac12 \sum_{n=1}\left[(\partial^\mu Z^4_n)^2+\frac{n^2}{R^2}(Z_n)^2-2\frac{n}{R}(\partial^\mu
Z^4_n)Z_{n,\mu}\right]\nonr\\
&&-\frac{1}{2\eta}\sum_{n=0}\left[(\partial_\mu Z^\mu_n)^2+2\frac{n}{R}\partial_\mu Z^\mu_n
Z_n^4
+\frac{n^2}{R^2}(Z_n^4)^2\right]\nonr\\
&&+{(g v_0 )^2\over 8}\left[Z_{0,\mu}Z_0^{\mu}+\sum_{n=1}(Z_{n,\mu} Z_n^{\mu}-Z^{4}_nZ^{4}_n)\right]
-\frac12 \sum_{n=0} W^{\mu\nu+}_n W^-_{n,\mu\nu}\nonr\\
&&+ \sum_{n=1}\left[(\partial^\mu W^{4+}_n\partial_\mu W^{4-}_n)+\frac{n^2}{R^2}(W^+_{n\mu}W_n^{\mu-})
-\frac{n}{R}(\partial^\mu W^{4-}_nW_{n\mu}^+ +\partial^\mu W_n^{4+}W_{n\mu}^-)\right]\nonr\\
&&-\frac{1}{\xi}\sum_{n=0}\left[(\partial_\mu W^{\mu+}_n)(\partial_\mu W^{\mu-}_n)
+\frac{n}{R}(\partial_\mu W^{\mu+}_n W_n^{4-}+\partial_\mu W^{\mu-}_n W_n^{4+} )
+\frac{n^2}{R^2}(W_n^{4+}W_n^{4-})\right]\nonr\\
&&+{(g v_0 c)^2\over
4}\left[W_{0,\mu}^+W_0^{-,\mu}+\sum_{n=1}(W_{n,\mu}^+W_n^{-,\mu}-W^{4+}_nW^{4-}_n)\right]\nonr\\
&&+\frac12\sum_{n=0}\left[(\partial^\mu h_n^0)^2-\frac{n^2}{R^2}(h_n^0)^2\right]
+\frac12\sum_{n=0}\left[(\partial^\mu\chi_n^0)^2-\left(\eta \left({g v_0\over
2}\right)^2+\frac{n^2}{R^2}\right)(\chi_n^0)^2\right]\nonr\\
&&+ \sum_{n=0}\left[\partial^\mu h_n^+ \partial_\mu h_n^--\left(\xi \left({g v_0 c\over 2}\right)^2+\frac{n^2}{R^2}\right)h_n^+
h_n^-\right]+ \cdots
\eeqa

From the above expansion, the propagators can be read:
\begin{center}
\begin{picture}(400,215)(0,30)

 \DashLine(20,45)(80,45){5}
 \Text(50,55)[c]{$h^0_n$}
 \Text(130,45)[l]{${i \over p^2-( m_H^2+\frac{n^2}{R^2})}$}

\DashLine(20,75)(80,75){5}
 \Text(50,85)[c]{$h^+_n$}
 \Text(130,75)[l]{${i \over p^2-(\xi M_W^2+\frac{n^2}{R^2})}$}

\DashLine(20,105)(80,105){5}
 \Text(50,115)[c]{$\chi^0_n$}
 \Text(130,105)[l]{${i \over p^2-(\eta M_Z^2+\frac{n^2}{R^2})}$}

 \DashLine(220,45)(280,45){5}
 \Text(250,55)[c]{$P^4_n$}
 \Text(330,45)[l]{${i \over p^2-\frac{n^2}{\alpha R^2} }$}

\DashLine(220,75)(280,75){5}
 \Text(250,85)[c]{$Z^4_n$}
 \Text(330,75)[l]{${i \over p^2-( M_Z^2+\frac{n^2}{\eta R^2})}$}

\DashLine(220,105)(280,105){5}
 \Text(250,115)[c]{$W^{\pm 4}_n$}
 \Text(330,105)[l]{${i \over p^2-(M_W^2+\frac{n^2}{\xi R^2})}$}

\Photon(20,145)(80,145){3}{4.5}
\Text(20,137)[c]{$\mu$} \Text(80,137)[c]{$\nu$}
 \Text(50,155)[c]{$W^\pm_n$}
 \Text(130,145)[l]{${-i \over p^2-(M_W^2+\frac{n^2}{R^2})}
\left\{g^{\mu\nu}+{(\xi-1)p^\mu p^\nu \over p^2-\xi(M_W^2+
\frac{n^2}{R^2})}\right\}$}

 \Photon(20,185)(80,185){3}{4.5}
 \Text(20,177)[c]{$\mu$} \Text(80,177)[c]{$\nu$}
 \Text(50,195)[c]{$Z_n$}
 \Text(130,185)[l]{${-i \over p^2-(M_Z^2+\frac{n^2}{R^2})}
\left\{g^{\mu\nu}+{(\eta-1)p^\mu p^\nu \over p^2-\eta(M_Z^2+
\frac{n^2}{R^2})}\right\}$}

\Photon(20,225)(80,225){3}{4.5}
\Text(20,217)[c]{$\mu$} \Text(80,217)[c]{$\nu$}
 \Text(50,235)[c]{$P_n$}
 \Text(130,225)[l]{${-i \over p^2-\frac{n^2}{R^2}}
\left\{g^{\mu\nu}+{(\alpha-1)p^\mu p^\nu \over p^2-\alpha \frac{n^2}{R^2}}\right\}$}
\end{picture}
\end{center}
For the Feynman gauge used in this calculation we set $\alpha=\eta=\xi=1$
otherwise there appear the following mixing vertex
\begin{center} \begin{picture}(200,110)(0,0)
\Photon(30,20)(60,20){2}{3}
\Text(10,20)[c]{$W^{\pm\mu}_n$}
\Line(55,25)(65,15) \Line(65,25)(55,15)
\DashLine(60,20)(90,20){3}
\ArrowLine(76,20)(78,20) \Text(75,10)[c]{$p$}
\Text(110,20)[c]{$W^{\pm 4}_n$}
\Text(180,20)[l]{$-\frac{n}{R}\frac{1-\xi}{\xi} p^\mu$}

\Photon(30,60)(60,60){2}{3}
\Text(10,60)[c]{$Z^{\mu}_n$}
\Line(55,65)(65,55) \Line(65,65)(55,55)
\DashLine(60,60)(90,60){3}
\ArrowLine(76,60)(78,60) \Text(75,50)[c]{$p$}
\Text(110,60)[c]{$Z^{ 4}_n$}
\Text(180,60)[l]{$-\frac{n}{R}\frac{1-\eta}{\eta} p^\mu$}

\Photon(30,100)(60,100){2}{3}
\Text(10,100)[c]{$P^{\mu}_n$}
\Line(55,105)(65,95) \Line(65,105)(55,95)
\DashLine(60,100)(90,100){3}
\ArrowLine(76,100)(78,100) \Text(75,90)[c]{$p$}
\Text(110,100)[c]{$P^{4}_n$}
\Text(180,100)[l]{$-\frac{n}{R}\frac{1-\alpha}{\alpha} p^\mu$}

\end{picture} \end{center}

 The brane or zero mode fermion couplings in terms of mass eigenstates are summarized in
 the following figure,
 where $\hat{R}/\hat{L}=\frac12(1\pm\gamma^5)$
 and summation is understood for the repeating indices.
\begin{center}
\begin{picture}(400,400)(15,-195)
\ArrowLine(20,10)(40,35) \Text(10,10)[c]{$l_j$}
\ArrowLine(40,35)(20,60) \Text(12,58)[c]{$l_i$}
\Photon(40,35)(70,35){3}{2.5} \Text(60,50)[c]{$A_n^\mu$}
\Text(100,20)[lb]{$ -i  e K_n\exp[-\frac{n^2\sigma^2}{4R^2}] \gamma^\mu
\left[ (V^L)^\dag_{ik}\cos\frac{n y_k^L}{R} V_{kj}^L \hat{L}
+(V^R)^\dag_{ik}\cos\frac{n y_k^R}{R} V_{kj}^R \hat{R} \right]$}

\ArrowLine(20,80)(40,105) \Text(10,80)[c]{$l_j$}
\ArrowLine(40,105)(20,130) \Text(12,128)[c]{$l_i$}
\Photon(40,105)(70,105){3}{2.5} \Text(60,120)[c]{$Z_n^\mu$}
\Text(100,90)[lb]{$ \frac{i g_2}{\cos\theta_W}K_n \exp[-\frac{n^2\sigma^2}{4R^2}]\gamma^\mu\!
\left[ g_L(V^L)^\dag_{ik}\!\cos\!\frac{n y_k^L}{R} V_{kj}^L\hat{L}
+g_R(V^R)^\dag_{ik}\!\cos\!\frac{n y_k^R}{R} V_{kj}^R \hat{R} \right]$}

\ArrowLine(20,150)(40,175) \Text(10,150)[c]{$l_j$}
\ArrowLine(40,175)(20,200) \Text(12,198)[c]{$\nu_i$}
\Photon(40,175)(70,175){3}{2.5} \Text(60,190)[c]{$W_n^{+\mu}$}
\Text(100,160)[lb]{$ \frac{i g_2}{\sqrt{2}} K_n\exp[ -\frac{n^2\sigma^2}{4R^2}] \gamma^\mu
\left[ (V^L)^\dag_{ik}\cos\frac{n y_k^L}{R} V_{kj}^L \hat{L} \right]$}

\ArrowLine(40,-35)(20,-10) \Text(10,-10)[c]{$l_i$}
\ArrowLine(20,-60)(40,-35) \Text(12,-58)[c]{$l_j$}
\DashLine(40,-35)(70,-35){4} \Text(60,-50)[c]{$h^0_n$}
\Text(100,-35)[lb]{$ -i\frac{K_n}{\sqrt{2}}\left[
\frac12(\lambda^{LR}_{ij,n}+\lambda^{LR *}_{ji,n} )
+\frac12(\lambda^{LR}_{ij,n}-\lambda^{LR *}_{ji,n})\gamma^5\right]
$}

\ArrowLine(40,-105)(20,-80) \Text(10,-80)[c]{$l_i$}
\ArrowLine(20,-130)(40,-105) \Text(12,-128)[c]{$l_j$}
\DashLine(40,-105)(70,-105){4} \Text(60,-120)[c]{$\chi^0_n$}
\Text(100,-105)[lb]{$   \frac{K_n}{\sqrt{2}}\left[
\frac12(\lambda^{LR}_{ij,n}-\lambda^{LR *}_{ji,n} )
+\frac12(\lambda^{LR}_{ij,n}+\lambda^{LR *}_{ji,n})\gamma^5\right]$}

\ArrowLine(40,-175)(20,-150) \Text(10,-150)[c]{$\nu_i$}
\ArrowLine(20,-200)(40,-175) \Text(12,-198)[c]{$l_j$}
\DashLine(40,-175)(70,-175){4} \Text(60,-190)[c]{$h^+_n$}
\ArrowLine(57,-175)(55,-175)
\Text(100,-175)[lb]{$ -i\sqrt{2} K_n \lambda^{LR}_{ij,n} \hat{R}$}
\end{picture}
\end{center}
where $K_n= \delta_{n,0}+\sqrt{2}(1-\delta_{n,0})$
 and $\lambda^{LR}_{ij,n}= \exp[ -\frac{n^2\sigma^2}{4R^2}]
\left[ (V^L)^\dag_{ik} f_{kl}\exp[-{(\triangle_{kl}^{LR})^2\over 4\sigma^2}]
\cos\frac{n \bar{y}_{kl}^{LR}}{R} V_{lj}^R  \right]$.
One can check that when $n=0$ they reduce to the usual flavor diagonal SM
couplings. Note that in general for $n\geq1$, $\lambda^{LR}_{ij,n}\neq\lambda^{LR
*}_{ji,n}$.
\newpage
\bibliographystyle{unsrt}

\begin{thebibliography}{99}
\bibitem{ADD}
N.~Arkani-Hamed, S.~Dimopoulos and G.~Dvali,
Phys.\ Lett.\  {\bf B429}, 263 (1998)
[hep-ph/9803315]

\bibitem{Anton}
I.~Antoniadis,
Phys.\ Lett.\  {\bf B246}, 377 (1990).

\bibitem{JL}
J.~Lykken,
Phys.\ Rev.\ {\bf D54}, 3693 (1996).

\bibitem{AS}
N.~Arkani-Hamed and M.~Schmaltz,
Phys.\ Rev.\  {\bf D61}, 033005 (2000)
[hep-ph/9903417]\\
N.~Arkani-Hamed, Y.~Grossman and M.~Schmaltz, Phys.\ Rev.\ {\bf D61}, 115004 (2000)
[hep-ph/9909411].

\bibitem{MS}
E.A.~Mirabelli and M. Schmaltz,
Phys.\ Rev.\ {\bf D61}, 113011 (2000)

\bibitem{Quiros}
A.~Delgado, A.~Pomarol and M.~Quiros,
 J.~High\ Energy\ Phys.\ {\bf 0001},030(2000)

\bibitem{Branco}
G.C.~ Branco, A.~de Gouvea and M.N.~Rebelo, Phys.\ Lett.\ {\bf B506}, 115 (2000)

\bibitem{bnu}
N.~Akani-Hamed, S.~Dimopoulos, G.~Dvali, and J.~March-Russell, hepph 9811448 (1998)\\
K.R.~Dienes,E.~Dudas, and T.~Gherghetta, Nucl.\ Phys.\ {\bf B577}, 25 (1999)

\bibitem{bnu1}
J.-M.~Fr\a`ere, M.V.~Libanov, and S.V. Troitsky, JHEP \ {\bf 0111}, 025 (2001)\\
C.S.~Lam, and J.N.~Ng, Phys.\ Rev \ {\bf 64D}, 113006 (2001)\\
C.S.~Lam ,{\it ibid},  {\bf 65D}, 053009 (2002)\\
K.R.~Dienes and I.~Sarcevic, Phys. \ Lett. {\bf B500}, 133 (2001)

\bibitem{bnup}
A.E.~Faraggi and M.E.~Pospelov, Phys. \ Lett. {\bf B458} 237 (1999)\\
G.C.~McLaughlin and J.N.~Ng, {\it ibid} {\bf B470}, 157 (1999); {\it ibid} {\bf B493}, 88 (2000)\\
G.C.~McLaughlin and J.N.~Ng, Phys.\ Rev. {\bf D63} 053002 (2001)

\bibitem{Tait}
D.E.~Kaplan and T.M.P.~Tait, JHEP {\bf 0111}, 051 (2001)

\bibitem{das}
F.~del Aguila and J.~Santiago, JHEP {\bf 0203}, 010 (2002)

\bibitem{univ}
T.~W.~Appelquist, J.~R.~Primack and H.~R.~Quinn,
Phys.\ Rev.\ {\bf D7}, 2998(1973);\\
W.~J.~Marciano and A.~Sirlin,
Phys.\ Rev.\ {\bf D8}, 3612(1973).

\bibitem{Baren}
G.~Barenboim, G.C.~Branco, A.~de Gouvea and M.N. Rebelo, \ Phys.\ Rev {\bf D64} 073005 (2001)


 \bibitem{trapf}
V.A.~Rubakov and M.E~ Shaposhnikov, \ Phys.\ Lett.\ {\bf B125}, 136 (1983)\\
R.~Jackiw and C.~Rebbi,\ Phys. \ Rev.\ {\bf D13}, 3398 (1976)

\bibitem{DS}
G.~Dvali and M.~Shifman, \ Phys. \ Lett.\ {\bf B475}, 295 (2000)

\bibitem{acd}
C.D.~Carone, Phys.\ Rev {\bf D61}, 015008 (2000)\\
T.~Appelquist,H.C.~Cheng, and B.~Dobrescu, {\it ibid} {\bf D64} 035002 (2001)

\bibitem{KKG}
W.~J.~Marciano, \ Phys.\ Rev.\ {\bf D60}, 093006 (1999)\\
M.~Masip and A.~Pomarol, \ Phys.\ Rev {\bf D60} 096005 (1999)

\bibitem{PDG}
D.E. Groom et al, The Euro.\ Phys.\ J.{\bf C15}, 1(2000)

\bibitem{ssum}
We use the formular $\sum^{\infty}_{n=1}\frac{\cos (n\pi \triangle)}{n^2}=\frac{\pi^2}{6}\left[
1-3|\triangle|+\frac{3}{2}\triangle ^2\right]$ and has a periodicity of $\triangle= \mathrm{mod\ } 2$.

\bibitem{Ross}
L. Ib$\acute{a}\tilde{n}$ez and G.G. Ross, Phys. \ Lett.\ {\bf B332}, 100(1994).

\bibitem{Jap}
H. Nishiura, K. Matsuda, T. Kikuchi and T. Fukuyama,
[hep-ph/0202189].

\bibitem{MAMC}
F.~Feinberg and S.~Weinberg,\ Phys.\ Rev.\ Lett.\ {\bf
6}, 381(1961);\\
K.~Horikawa and K.~Sasaki,\ Phys.\ Rev.\ {\bf D 53}, 560(1996)

\bibitem{MAMC_exp}
L.~Willmann et al, Phys.\ Rev.\ Lett.\ {\bf82}, 49(1999).


\bibitem{KLN}
T.~Kinoshita, J.\ Math.\ Phys.\ {\bf 3}, 650(1962)\\
T.~D.~Lee and M.~Nauenberg, Phys.\ Rev.\ {\bf 133}, B1549(1964).

\bibitem{LHC}
F.~Gianotti and M.~P.~Altarelli,
Nucl.\ Phys.\ {\bf B}(Proc. Suppl.){\bf 89},177(2000).

\end{thebibliography}

\end{document}